\documentclass[referee]{raa}            

\usepackage{graphicx,bm}             
\begin{document}

   \title{Precession Effects on Liquid Planetary Core
}

   \volnopage{Vol.0 (200x) No.0, 000--000}      
   \setcounter{page}{1}          

   \author{Min Liu
      \inst{1,2}
   \and Ligang Li
      \inst{1}
   }

   \institute{Shanghai Astronomical Observatory, CAS,
             Shanghai 200030, China; \\
        \and
             University of Chinese Academy of Sciences, Beijing 100049, China;\\
             {\it mliu@shao.ac.cn}
   }

   \date{Received~~2015 Oct. day; accepted~~2015~~month day}

\abstract{
Motivated by the desire to understand the rich dynamics of precessionally driven flow in the liquid planetary core,
we investigate, through numerical simulations, the precessing fluid motion in a rotating cylindrical annulus
which possesses slow precession simultaneously.
The same problem has been studied extensively in cylinders where the precessing flow is characterized by three key parameters: the Ekman number $E$, the Poincar$\acute{\mathrm e}$ number $Po$ and the radius-height aspect ratio $\Gamma$.
While in an annulus, there is another parameter, the inner-radius-height aspect ratio $\Upsilon$, which also plays an important role in controlling the structure and evolution of the flow.
By decomposing the nonlinear solution into a set of inertial modes, we demonstrate the
properties of both weakly and moderately precessing flows.
It is found that, when the precessional force is weak, the flow is stable with a constant amplitude of kinetic energy. As the precessional force increases, our simulation suggests that the nonlinear interaction between the boundary effects and the inertial modes can trigger more turbulence, introducing a transitional regime of rich dynamics to disordered flow. The inertial mode $\bm u_{111}$, followed by $\bm u_{113}$ or $\bm u_{112}$, always dominates the precessing flow when $0.001\leq Po\leq 0.05$, ranging from weak to moderate precession. Moreover, the precessing flow in an annulus shows more stability than in a cylinder which is likely to be caused by the effect of the inner boundary that restricts the growth of resonant and non-resonant inertial modes.
Furthermore, the mechanism of triadic resonance is not found in the transitional regime from the laminar to disordered flow.
}

   \authorrunning{M. Liu \& L.-G. Li }            
   \titlerunning{Precession Effects On Liquid Planetary Core}  

   \maketitle

%
%
\section{Introduction}           
\label{sect:intro}
With significant applications in various research fields, the problem of precession-driven flow has been investigated for a long time by theoretical, experimental and numerical studies (Wood~\cite{Wood1966}; Gans~\cite{Gans1970}; Kobine~\cite{Kobine1995}; Tilgner \& Busse~\cite{Tilgner2001}; Zhang et al.~\cite{Zhang2010}).
In the planetary fluid dynamics, it is suggested that the precessing flow, instead of thermal or compositional convection, might be a candidate to generate geomagnetic field (Malkus~\cite{Malkus1994}; Tilgner~\cite{Tilgner2007}). In aerospace industry, the precession-driven flow in the fuel tank may also induce the hazardous instability of a spinning spacecraft (Bao et al.~\cite{Bao1997}). And in the laboratory experiments of rotating fluids, the precessing flow always arises from the Earth's rotation if it is not parallel to the rotation of the fluid container (Boisson et al.~\cite{Boisson2012}).

Extensive studies have been done on this problem in spherical, spheroidal, cylindrical and annular geometries. In a sphere, spherical shell or weakly deformed spheroid, the precession induces a flow of constant vorticity in the precessing frame which can be described by an inertial wave mode with azimuthal wave number $m=1$ in the mantle frame of reference (Busse~\cite{Busse1968}; Noir et al.~\cite{Noir2001}; Zhang et al.~\cite{Zhangetal2010}; Kida~\cite{Kida2011}). Meanwhile, the geostrophic flow develops from the inner and/or outer boundary layers. When the precession rate is large enough, the large-scale precessional flow becomes unstable and leads to a growth and collapse of small scales. The key difference between annulus and sphere is that annulus has more geometrical parameters to tune, which reveals richer dynamics in this problem.

However, in cylindrical geometry, the precession-driven flow does not reduce to a uniform vorticity. The flow can be described by a superposition of cylindrical inertial modes and viscous boundary effects (Liao \& Zhang~\cite{Liao2012}). When the precession rate is small, a single inertial mode with simple spatial structure and time-independent amplitude is directly driven by the precessional force at resonance and dominates the precessing flow (laminar flow). As precession rate increases, the simple spatial structure breaks into complex and the velocity amplitude becomes time-dependent because of the nonlinear interactions between inertial modes and viscous boundary layers (disordered flow). This could be one of the mechanisms to explain the breakdown of laminar flow and the transitional dynamics to disordered flow when Poincar$\acute{\mathrm e}$ force increases (Kong et al.~\cite{Kong2015}).

There is also a different mechanism of \emph{triadic resonance} to explain how  laminar flow transits to disordered flow in precessing cylinders (Kobine~\cite{Kobine1996}; Lagrange et al.~\cite{Lagrange2008}), in which the breakdown of laminar flow is rooted from the nonlinear interactions among three inertial modes. The primary inertial mode (e.g., $\bm u_{mnk}$: the flow velocity with azimuthal, axial and radial wave number $m, n, k$, respectively), which is directly driven by the Poincar$\acute{\mathrm e}$ force (Poincar$\acute{\mathrm e}$ mode), can be reinforced by the nonlinear interaction between other inertial modes (e.g., $\bm u_{\tilde{m}\tilde{n}\tilde{k}}$ and $\bm u_{\hat{m}\hat{n}\hat{k}}$). At the same time , any two inertial modes involving the Poincar$\acute{\mathrm e}$ mode (e.g., $\bm u_{mnk}$ and $\bm u_{\hat{m}\hat{n}\hat{k}}$) can induce the other inertial mode (e.g., $\bm u_{\tilde{m}\tilde{n}\tilde{k}}$). When a triadic resonance occurs among the three modes of $\bm{u}_{mnk}, \bm{u}_{\tilde{m}\tilde{n}\tilde{k}}$ and $\bm{u}_{\hat{m}\hat{n}\hat{k}}$, the following parametric conditions must be satisfied,
   \begin{equation}
    |\tilde{m}-\hat{m}|=m,~~~~~|\tilde{n}-\hat{n}|=n,~~~~~|\tilde{\omega}-\hat{\omega}|=\omega,
    \label{triadic}
    \end{equation}
where $\omega$ refers to the angular frequency of corresponding mode.

Since planets generally have a solid core, cylindrical annulus might be a better geometry than cylinder to simulate the fluid dynamics in the equatorial region.  In addition, there is another advantages of annular geometry attracting more research attention that it is readily realizable in both laboratory experiments and numerical simulations where there is no singularity in the mathematical formulas in cylindrical coordinates.

In a laboratory experiment, Lin et al. (\cite{Lin2014}) reported that the triadic resonance exists in the precessing flow in an \emph{annulus}, but the numerical studies of Kong et al. (\cite{Kong2015}) and Jiang et al. (\cite{Jiang2015}) did not find any evidence that indicates the triadic resonance taking place in \emph{cylinders} with several different radius-height ratios. The question then arises: why are there different mechanisms to explain how the laminar flow becomes disordered as Poincar$\acute{\mathrm e}$ force increases in two similar geometries? Motivated by this, we will re-investigate the precessing flow in an annulus via numerical simulations and identify whether the triadic resonance does exist.

In the rest of the paper,  we will present the mathematical formulation of the problem in Section \ref{sect:maths} and describe the numerical method in Section \ref{sect:diff}. The results are provided in Section \ref{sect:result}, and a brief summary and discussion are supplied in Section \ref{sect:discuss}.

\section{Mathematical formulation}
\label{sect:maths}
\begin{figure}
   \centering
   \includegraphics[width=0.382\textwidth]{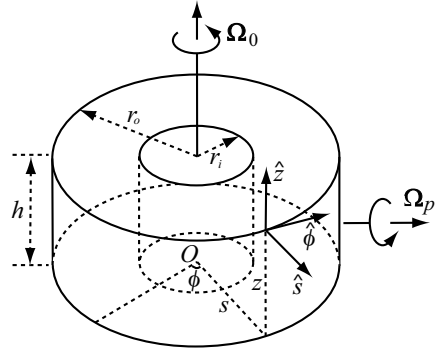}
   \caption{Geometry of a precessing cylindrical annulus with height $h=1$, inner radius $r_\mathrm{i}=0.269$ and outer radius $r_\mathrm{o}=1$. Cylindrical polar coordinates $(s,\phi,z)$ is fitted to the annulus while the precession angular velocity $\bm{\Omega}_p$ is fixed in space and perpendicular to the rotation angular velocity $\bm{\Omega}_0$.}
   \label{geometry}
\end{figure}

The mathematical model considers a homogeneous, incompressible fluid with constant viscosity $\nu$ and constant density $\rho$ fully filled in a cylindrical annulus of length $h$, inner radius $r_\mathrm{i}$ and outer radius $r_\mathrm{o}$. For convenience, the cylindrical polar coordinates $(s,\phi,z)$ with the corresponding unit vectors $(\hat{s},\hat{\phi},\hat{z})$ are adopted, where $s=0$ is at the symmetry axis of the annulus, $z=0$ is at the bottom (see Fig. \ref{geometry}). The cylindrical annulus rotates with an angular velocity $\bm{\Omega}_0=\Omega_0\hat{z}$ about its axis of symmetry, where $\Omega_0$ is a constant. Meanwhile, the annulus also undergoes slow precession with an angular velocity $\bm{\Omega}_p$ which is fixed in the inertial reference frame and perpendicular to $\bm{\Omega}_0$. We set $r_\mathrm{i}=0.269, r_\mathrm{o}=1$ and $h=1$, which are exactly the same as those of the experimental study by Lin et al. (\cite{Lin2014}) in order to identify the existence of triadic resonance.

In the $(s,\phi,z)$ coordinates fixed in the rotating annulus, the angular velocity of precession $\bm{\Omega}_p$ is time-dependent,
\begin{equation}\label{eq1}
    \bm{\Omega}_p=|\Omega_p|\,\left[\hat{s}\cos(\phi+\Omega_0 t)-\hat{\phi}\sin(\phi+\Omega_0 t)\right],
\end{equation}
where $|\Omega_p|$ is the positive constant amplitude. Then the governing equations of precessionally driven flow would be
\begin{eqnarray}
  \nonumber&{\partial\bm{u}}/{\partial t}+\bm{u}\cdot\nabla\bm{u}
  +2\left\{\hat{z}\Omega_0+|\Omega_p|\left[\hat{s}\cos(\phi+\Omega_0 t)-
  \hat{\phi}\sin(\phi+\Omega_0 t)\right]\right\}\times \bm{u}&\label{eqn_motion}\\
  &= -(1/\rho)\,\nabla p+\nu\nabla^2 \bm{u}-
  2\hat{z}|\Omega_p|\Omega_0s\cos(\phi+\Omega_0 t),& \\
  &\nabla\cdot \bm{u} = 0, &
\end{eqnarray}
where $\bm u$ and $p$ are the velocity vector and reduced pressure to be solved, respectively. The last term on the right hand side of Equation~(\ref{eqn_motion}) is the  driving force (Poincar$\acute{\mathrm e}$ force) to excite the precessing flow.

By using $h$, $1/\Omega_0$ and $\rho h^2\Omega_0^2$ as the scales of length, time and pressure, respectively, the non-dimensional equations are obtained as
\begin{eqnarray}
  \nonumber&{\partial\bm{u}}/{\partial t}+\bm{u}\cdot\nabla\bm{u}
  +2\left\{\hat{z}+Po\left[\hat{s}\cos(\phi+ t)-
  \hat{\phi}\sin(\phi+ t)\right]\right\}\times \bm{u}&\\
  &= -\nabla p+E\nabla^2 \bm{u}-
  2\hat{z}sPo\cos(\phi+ t),& \label{eqn_dimensionless1}\\
 & \nabla\cdot \bm{u} = 0,&\label{eqn_dimensionless2}
\end{eqnarray}
where the Poincar$\acute{\mathrm e}$ number $Po=|\bm{\Omega}_p/\bm{\Omega}_0|$ and the Ekman number $E=\nu/(|\bm{\Omega}_0|h^2)$.

In the rotating frame of reference, the flow on the boundary is non-slip, imposing
\begin{equation}\label{bound_a}
    u_s=u_\phi=u_z=0 \end{equation}
on the bounding surface of the annulus.
The initial condition is not important and can be arbitrary since the system will reach a  nonlinear equilibrium state.

The problem of precessing flow defined by (\ref{eqn_dimensionless1})-(\ref{eqn_dimensionless2}) subject to the boundary conditions (\ref{bound_a}) will be solved numerically by a finite difference method to reveal the evolution of precessing flow and check the existence of triadic resonance.

\section{Numerical method of 3-D finite difference}
\label{sect:diff}
Unlike a cylinder or sphere, there is no axial or polar singularity in an annulus when cylindrical polar coordinates are adopted, so it is very suitable to use finite difference method to solve this kind of problems.

The finite difference method used in our simulations is similar to that of Chan et al. (\cite{Chan2006}) which will be briefly described hereafter. In the spatial discretization, the annulus is divided uniformly in radial ($s$), azimuthal ($\phi$) and axial ($z$) coordinates (nonuniform grid also can be used in $s$ and/or $z$ coordinates).  A staggered grid is used to represent the unknown variable $\bm u$ and $p$ with which the radial component of velocity $u_s$ is located at $[(i+0.5)\Delta\phi,~r_\mathrm{i}+j\Delta s,~(k+0.5)\Delta z]$ (the nodal point of $s$ and midpoints of $\phi$ and $z$, here $k$ should not be confused with the radial wave number $k$ in the context), the azimuthal component $u_\phi$  is located at $[i\Delta\phi,~r_\mathrm{i}+(j+0.5)\Delta s,~(k+0.5)\Delta z]$ and the axial component $u_z$  at $[(i+0.5)\Delta\phi,~r_\mathrm{i}+(j+0.5)\Delta s,~k\Delta z]$, while the pressure $p$ is located at the center of the grid. All spatial operators are discretized by central difference scheme of a second order accuracy.

The approximate factorization method is used in temporal discretization (Dukowicz \& Dvinsky~\cite{Dukowicz1992}) by splitting the Crank-Nicolson scheme of the equations.  In this method, the time forwarding from $t_n$ to $t_{n+1}$ is split into two steps: a prediction and a correction step where velocity and pressure are decoupled, thus can be solved separately. The procedure is briefly described as follows.\\
\emph{The first step:} to predict a temporary velocity $\tilde{\bm u}$
\begin{equation}\label{time-discret}
    \left[I-({\Delta t}/{2})\,\mathcal{L}\right]\tilde{\bm u}=\left[I+({\Delta t}/{2})\,\mathcal{L}\right]\hat{\bm u}+\Delta t (\bm u\cdot\nabla\bm u)^{n+1/2}
\end{equation}
with
\begin{equation}
   \hat{\bm u}={\bm u}^n-({\Delta t}/{2})\,\mathcal{G}p^n,
\end{equation}
where $I$ is the unitary matrix, $\mathcal{L}$ is the discrete spatial operator referring to the linear terms in equation (\ref{eqn_dimensionless1}) and $\mathcal{G}$ is the discrete operator of gradient.
The nonlinear term at the intermediate time step in (\ref{time-discret}) is approximated explicitly by using the Adams-Bashford scheme of second order as
\begin{equation}
    (\bm u\cdot\nabla\bm u)^{n+1/2}=(3/2)~(\bm u\cdot\nabla\bm u)^n-(1/2)~(\bm u\cdot\nabla\bm u)^{n-1}+O(\Delta t^2).
\end{equation}

After solving the temporary velocity $\tilde{\bm u}$ in (\ref{time-discret}), the second step is to obtain $p^{n+1}$ from (\ref{correct1}) and ${\bm u}^{n+1}$ from (\ref{correct2}) as follows so that ${\bm u}^{n+1}$ satisfies the incompressible condition (\ref{eqn_dimensionless2}). It reads the following step. \\
\emph{The second step:} to solve the pressure and correct the temporary velocity
\begin{eqnarray}
&({\Delta t}/{2})\,\mathcal{DG}p^{n+1}=\mathcal{D}\tilde{\bm u},&\label{correct1}\\
&\bm u^{n+1}=\tilde{\bm u}-({\Delta t}/{2})\,\mathcal{G}p^{n+1},\label{correct2}&
\end{eqnarray}
where $\mathcal{D}$ is the discrete divergence operator. It is noted that this splitting scheme remains second order accuracy in time and space domains.

Finally, both steps will transform to a typical linear system $\bm{Ax}=\bm{b}$, which can be solved by iterative method. In this study, we take grid numbers in $(s,\phi,z)$ coordinates as (100,180,100) and time step $\Delta t=0.001$ respectively. The size of matrix $\bm{A}$ is about 5.4M$\times$5.4M (for $\tilde{\bm u}$, 1M$=10^6$) or 1.8M$\times$1.8M (for $p^{n+1}$), which requires parallel computation to solve the massive linear problems. The computational code has been verified to be correct and meet the second order precision.

As to parameters, when the Ekman number $E$ is sufficiently small such that the dynamics of precessing flow is dominated by the effects of rotation, the precise value of $E$ becomes physically and mathematically less significant. Since most laboratory experiments were performed in the range $10^{-5}\leq E\leq 10^{-4}$ (e.g., Kobine~\cite{Kobine1995}, Lin et al.~\cite{Lin2014}), we take $E=5\times10^{-5}$ in our numerical simulations.
 The Poincar$\acute{\mathrm e}$ number $Po$ is the key parameter to control the degree of nonlinearity of precessing flow, hence a few $Po$ values from 0.001 to 0.05 are adopted in the numerical simulations which extend the regime from weak precession ($Po/\sqrt{E}\ll 1$) to moderate precession ($1\leq Po/\sqrt{E}\leq 10$).

\section{Results}
\label{sect:result}
In the classical theory for \emph{cylindrical} geometry, there exist three controlling parameters that characterize the precessing flow: the Ekman number $E$, denoting the relative importance of the viscous force to the Coriolis force; the geometrical parameter $\Gamma$, representing the radius-height aspect ratio of the cylinder; and the Poincar$\acute{\mathrm e}$ number $Po$, measuring the magnitude of the precessional force. However, in an \emph{annulus}, the inner-radius-height aspect ratio $\Upsilon=r_{\mathrm{i}}/h$ also plays an important role not only in influencing the structure of precessing flow but also in controlling the time evolution of nonlinear precessing flow.
To elucidate the precise mechanism of instabilities of the precessional flow in an annulus, a good way is to expand the nonlinear flow into a complete set of inertial modes for a cylindrical annulus (the completeness of the inertial modes is partly proved by Cui et al.~\cite{Cui2014}). Following  Liao \& Zhang (\cite{Liao2012}), Kong et al. (\cite{Kong2015}) and Jiang et al. (\cite{Jiang2015}), the velocity field can be written as
\begin{eqnarray}
  {\bm u}(s,\phi,z,t) &=& \tilde{\bm u}+\sum_{k=1}^{K}A_{00k}(t)\bm u_{00k}(s)+\sum_{m=1}^{M}\sum_{k=1}^{K}\frac{1}{2}\left[A_{m0k}(t)\bm u_{m0k}(s,\phi)+\mathrm{c.c.}\right] \nonumber\\
   &&+\sum_{n=1}^{N}\sum_{k=1}^{K}\frac{1}{2}\left[A_{0nk}(t)\bm u_{0nk}(s,z)+\mathrm{c.c.}\right]\nonumber\\
   &&+\sum_{m=1}^{M}\sum_{n=1}^{N}\sum_{k=1}^{2K}\frac{1}{2}\left[A_{mnk}(t)\bm u_{mnk}(s,\phi,z)+\mathrm{c.c.}\right],\label{eqn_decomp}
\end{eqnarray}
 where c.c. represents the complex conjugate of the previous term. The multiple components of the precessing flow may include: the boundary layer flow $\tilde{\bm u}$; the axisymmetric, non-axisymmetric geostrophic mode $\bm u_{00k}(s)$, $\bm u_{m0k}(s,\phi)$ with $m\geq 1$; the axisymmetric oscillatory inertial mode $\bm u_{0nk}(s,z)$ with $n\geq 1$; and the non-axisymmetric inertial wave modes $\bm u_{mnk}(s,\phi,z)$  with $m\geq 1$ and $n\geq 1$. Note $k$ is always greater than and equal to 1. The precessing flow resulting from numerical simulations can be decomposed into the inertial eigenmodes to understand the mechanism of nonlinear interactions in an annulus.

\subsection{Structure of the precessing flow}
\begin{figure}
 \textbf{(a)}\\
   \centerline{\includegraphics[width=\textwidth]{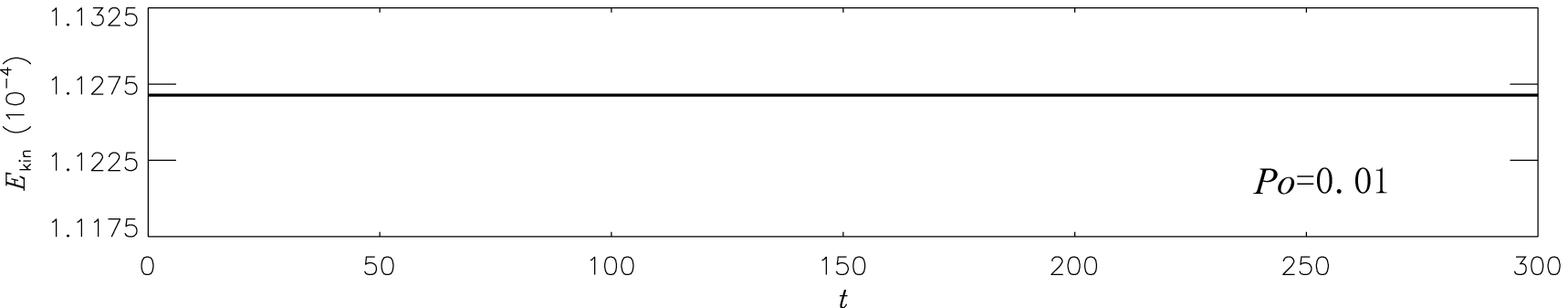}}\\
 \textbf{(b)}\\
   \centerline{\includegraphics[width=\textwidth]{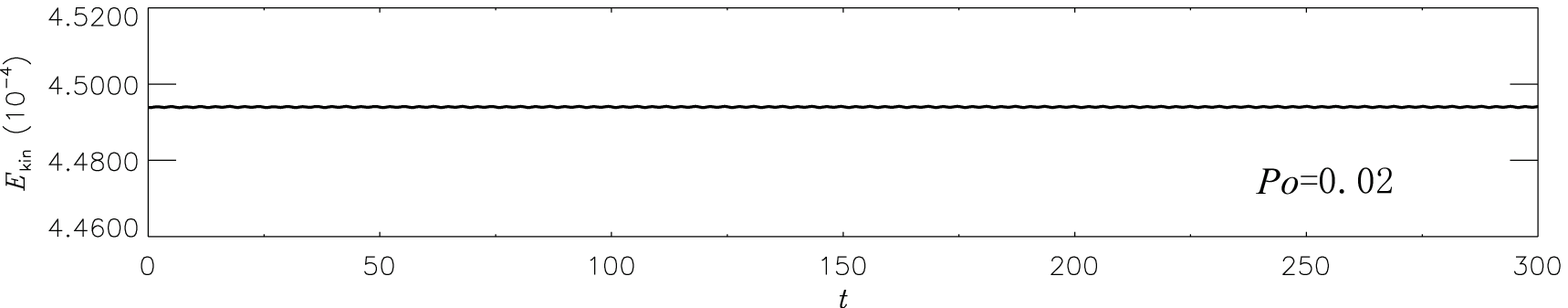}}\\
 \textbf{(c)}\\
   \centerline{\includegraphics[width=\textwidth]{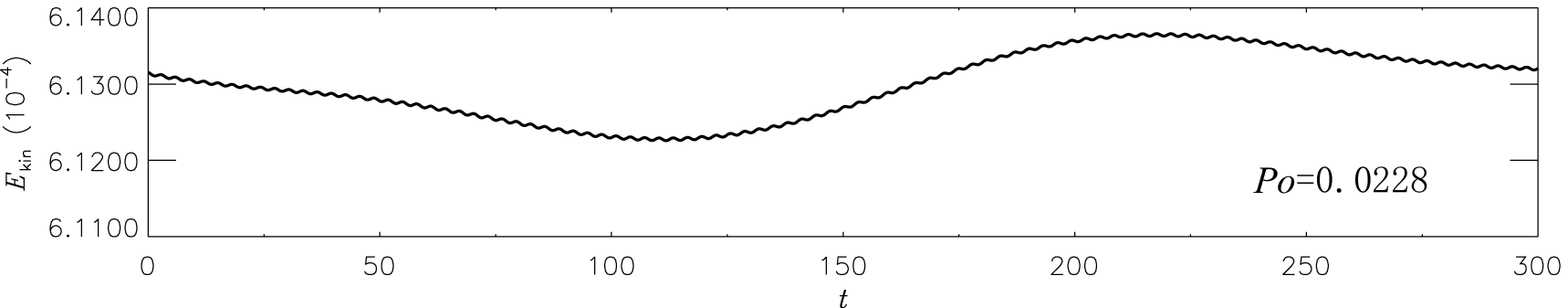}}\\
 \textbf{(d)}\\
   \centerline{\includegraphics[width=\textwidth]{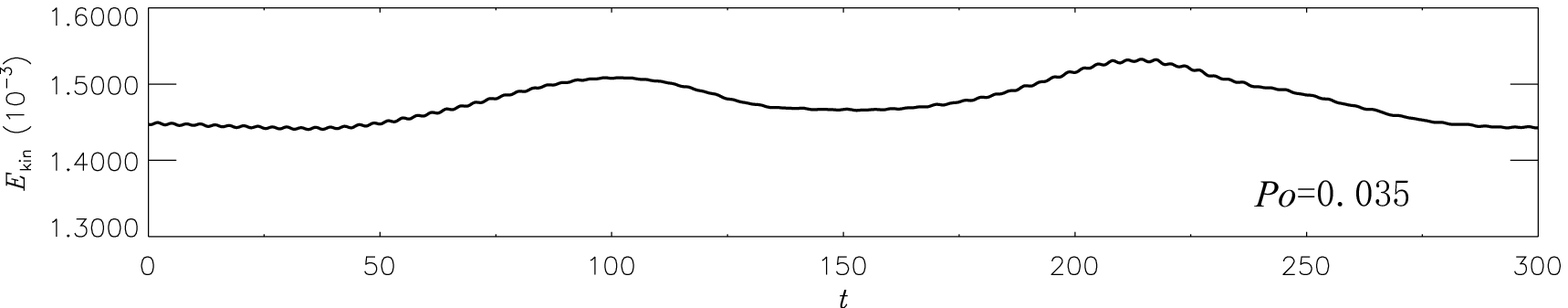}}\\
 \textbf{(e)}\\
   \centerline{\includegraphics[width=\textwidth]{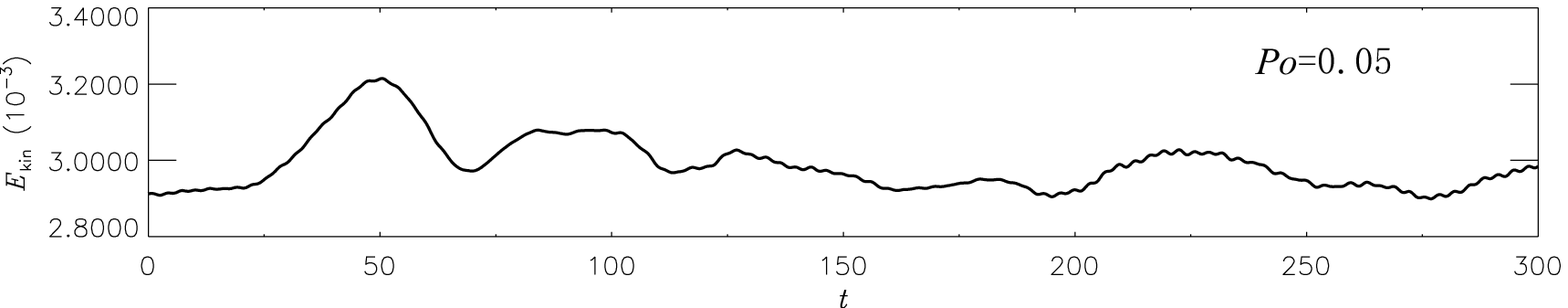}}
   \caption{Time-evolution of kinetic energy density $E_{\mathrm{kin}}$ of the precessing flow at $E=5\times 10^{-5}$ in the cylindrical annulus with $r_\mathrm{i}=0.269$ and $r_\mathrm{o}=1$ for (a) $Po=0.01$; (b) $Po=0.02$; (c) $Po=0.0228$; (d) $Po=0.035$ and (e) $Po=0.05$, respectively. }
   \label{figEkin}
\end{figure}

\begin{figure}
\begin{center}
    \begin{tabular}{rcl}
        \begin{minipage}{0.28\textwidth}
            \textbf{(a1)}\\
            \includegraphics[width=\textwidth]{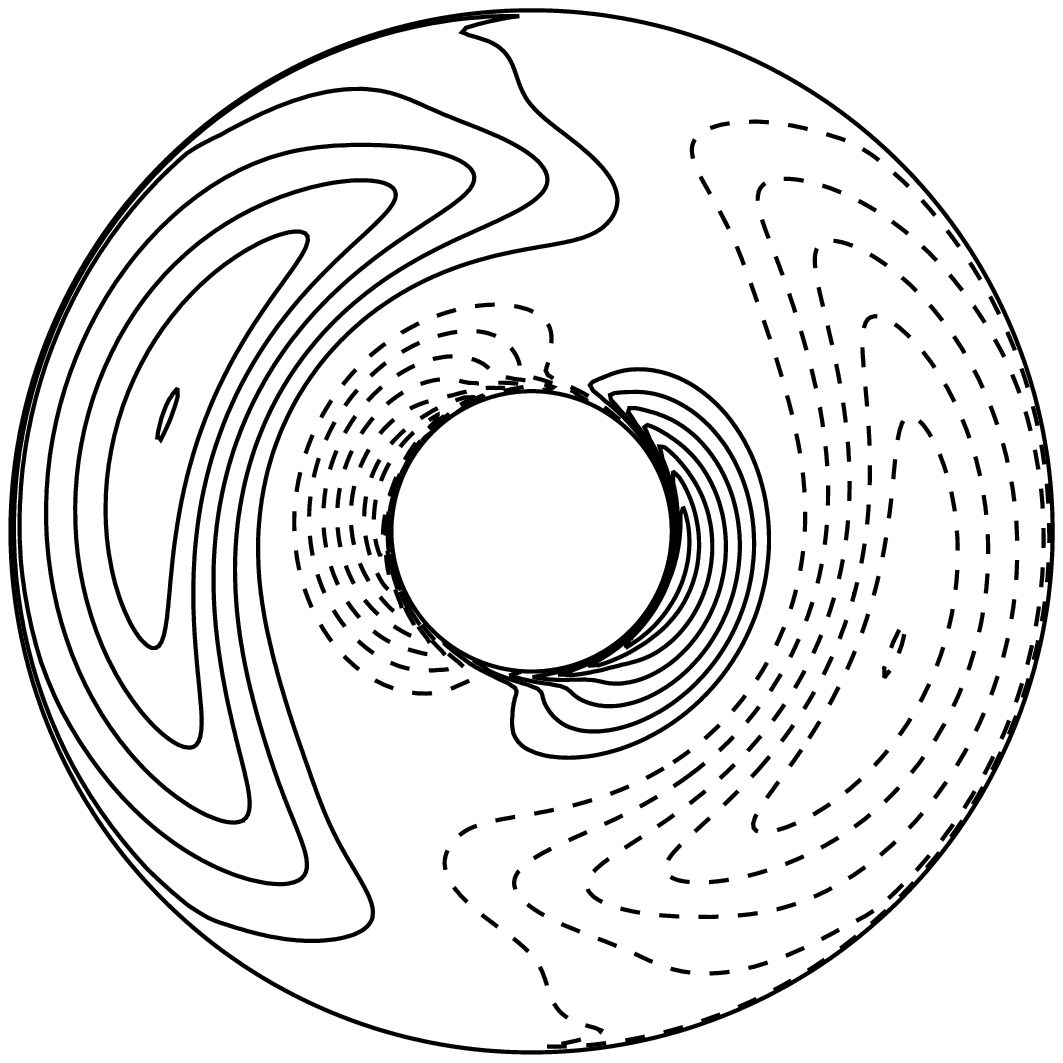}\\
            \vspace{-0.4cm}
            \textbf{(b1)}\\
            \includegraphics[width=\textwidth]{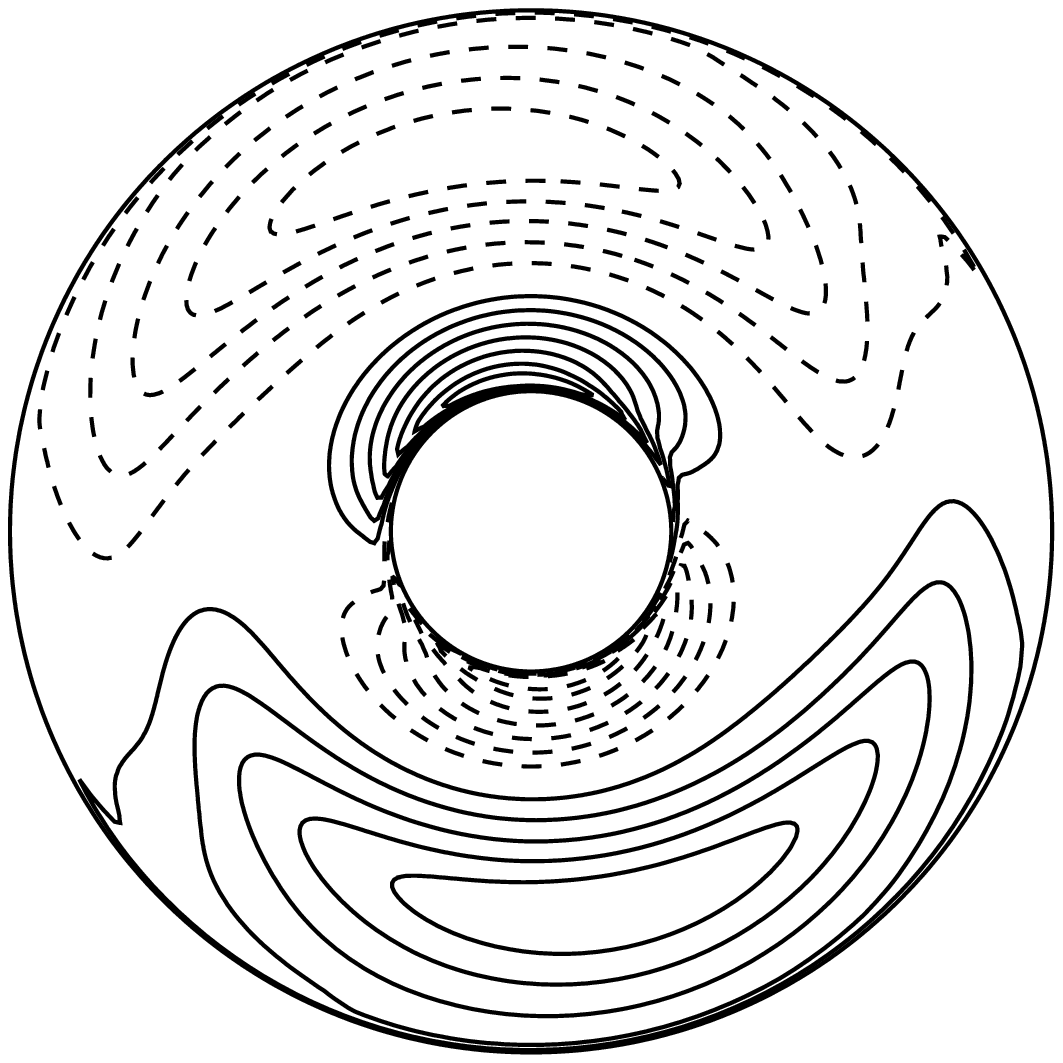}\\
            \vspace{-0.4cm}
            \textbf{(c1)}\\
            \includegraphics[width=\textwidth]{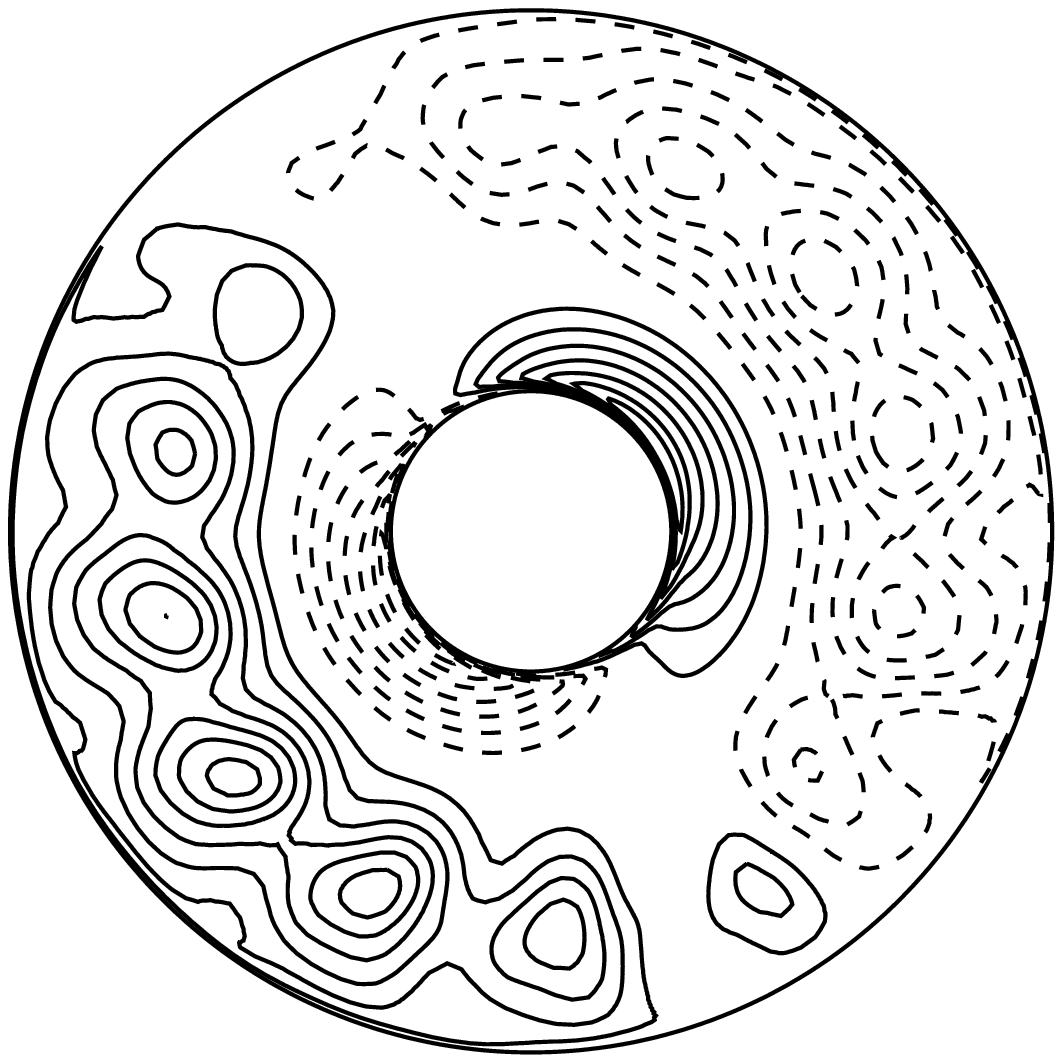}\\
            \vspace{-0.4cm}
            \textbf{(d1)}\\
            \includegraphics[width=\textwidth]{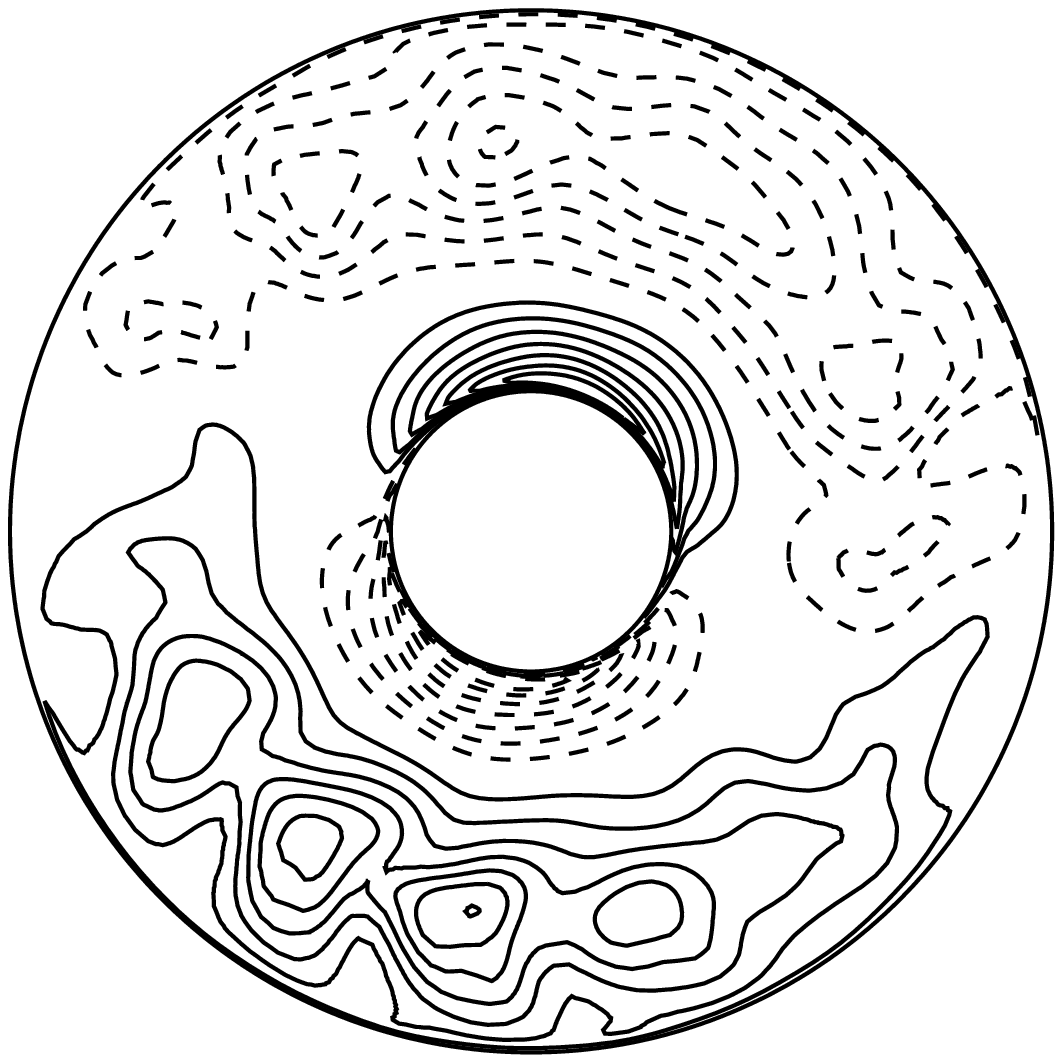}\\
            \vspace{-0.4cm}
            \textbf{(e1)}\\
            \includegraphics[width=\textwidth]{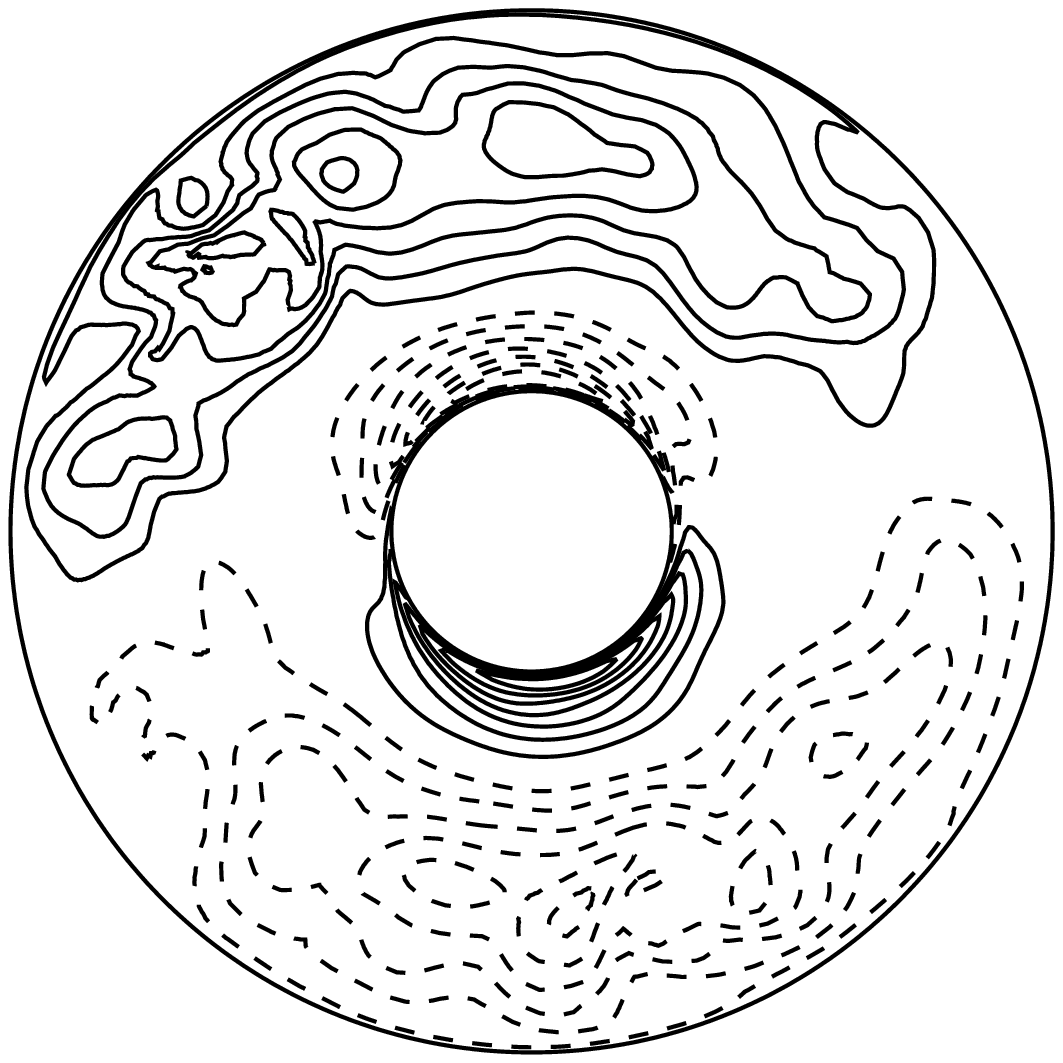}
        \end{minipage}
        ~~~
        \begin{minipage}{0.28\textwidth}
            \textbf{(a2)}\\
            \includegraphics[width=\textwidth]{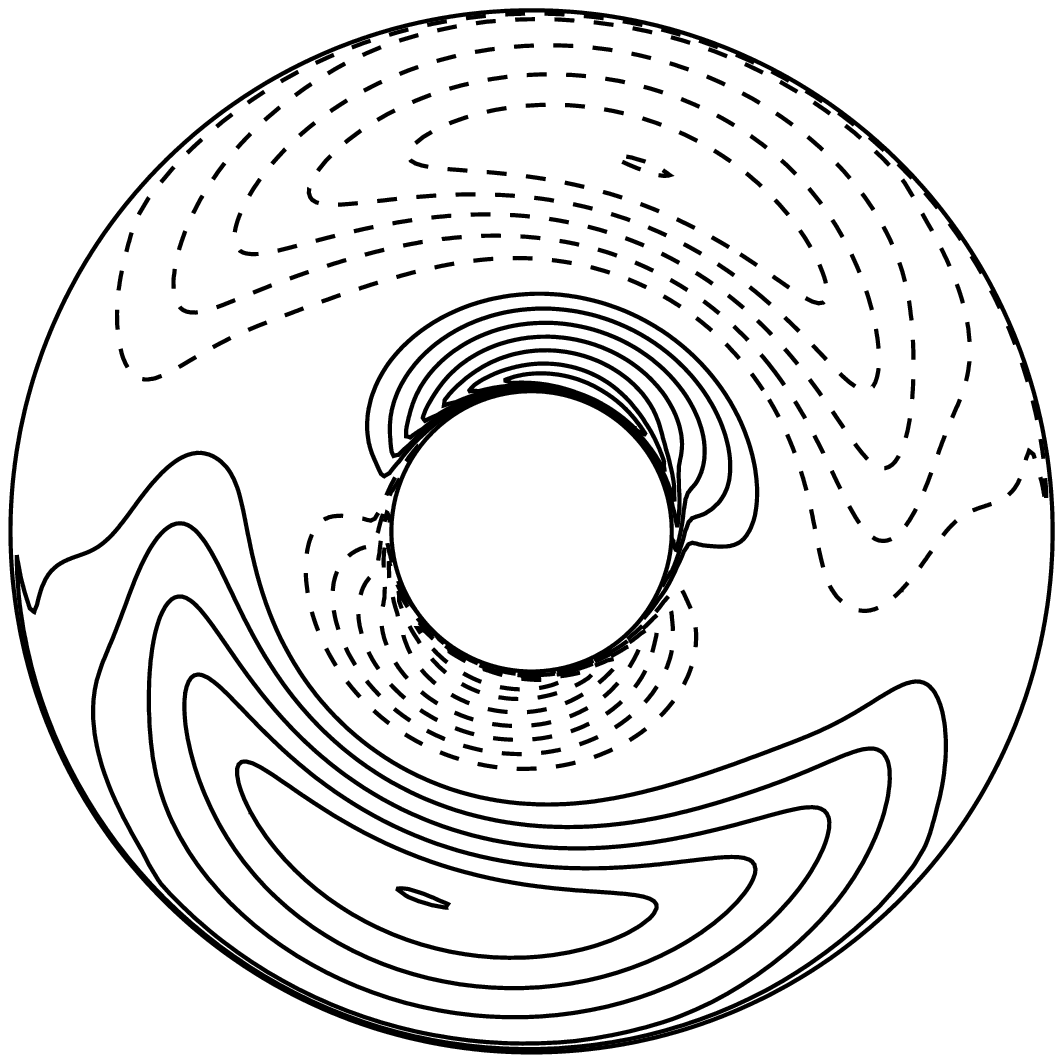}\\
            \vspace{-0.4cm}
            \textbf{(b2)}\\
            \includegraphics[width=\textwidth]{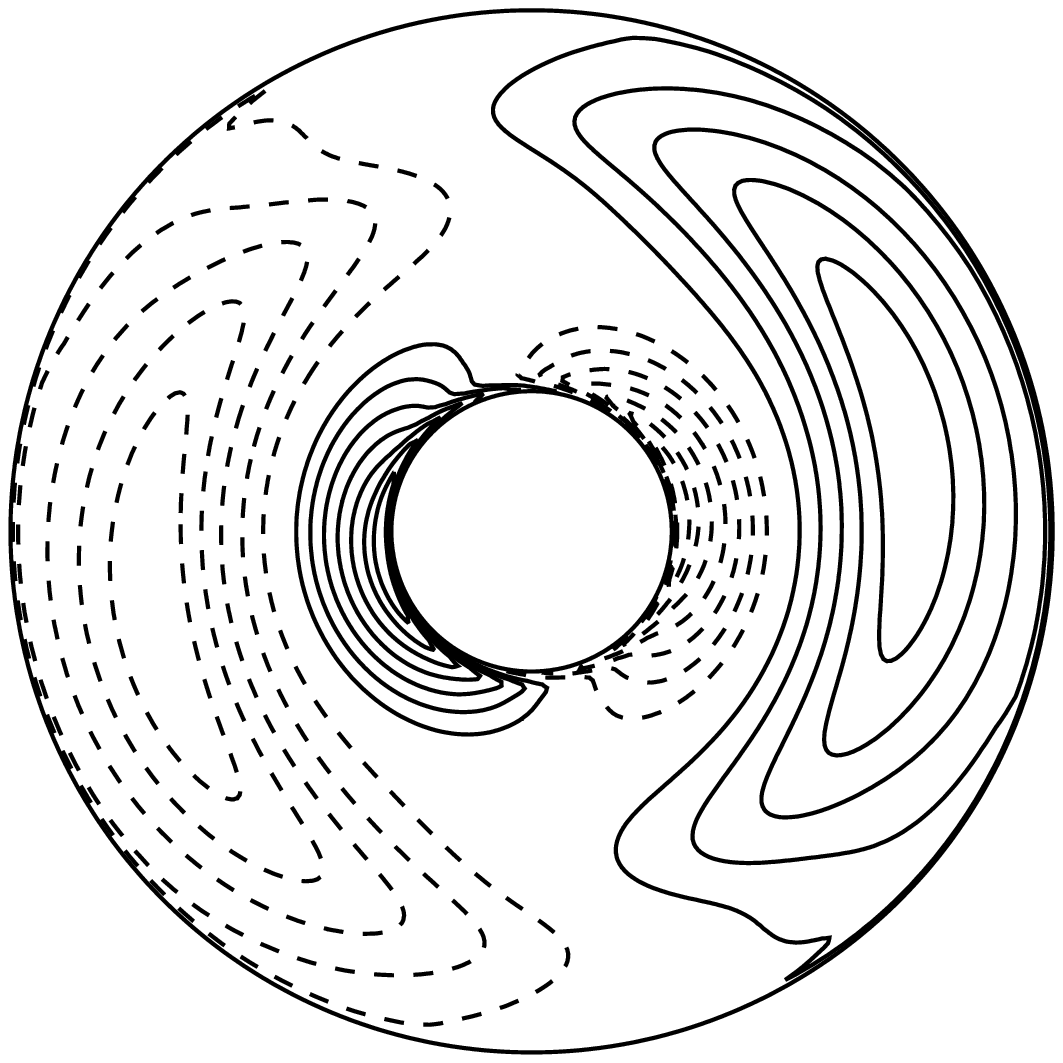}\\
            \vspace{-0.4cm}
            \textbf{(c2)}\\
            \includegraphics[width=\textwidth]{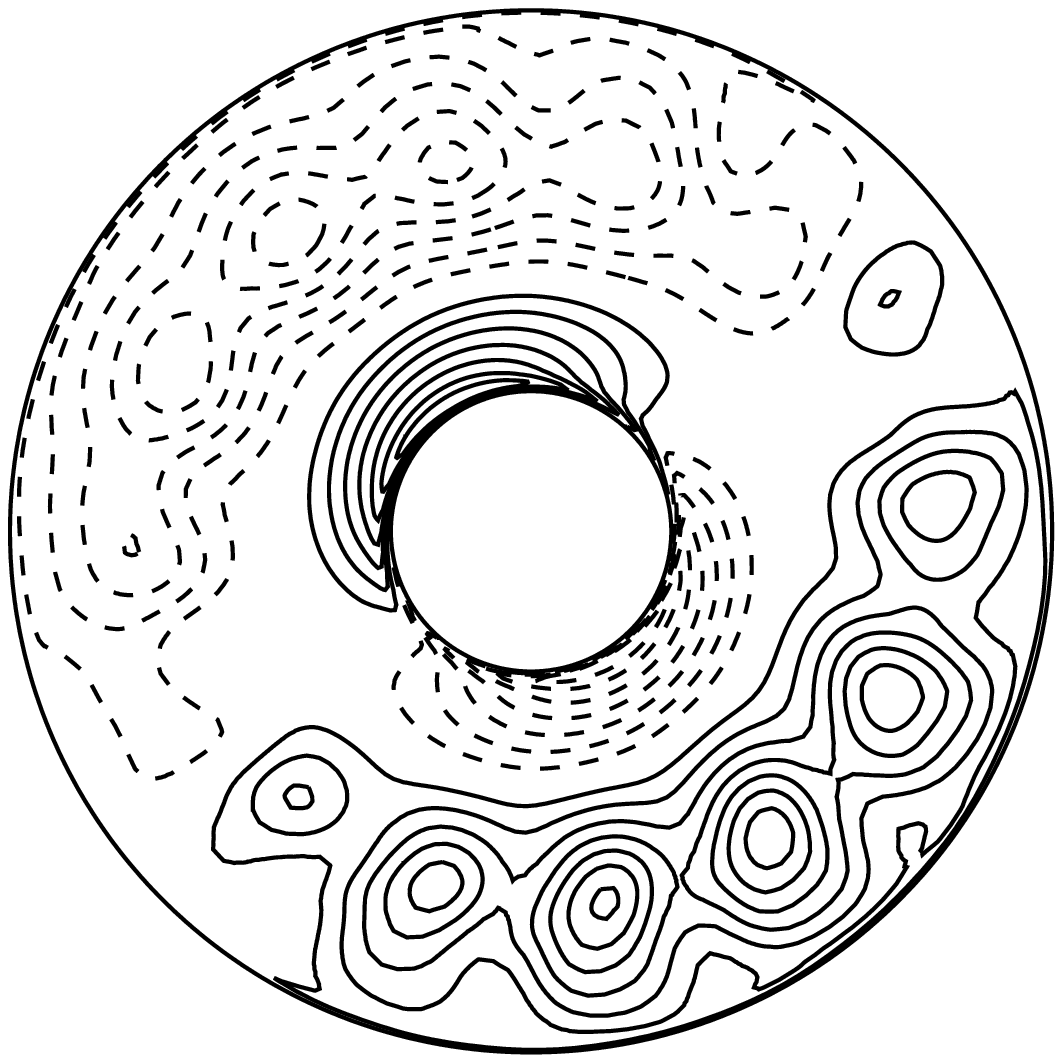}\\
            \vspace{-0.4cm}
            \textbf{(d2)}\\
            \includegraphics[width=\textwidth]{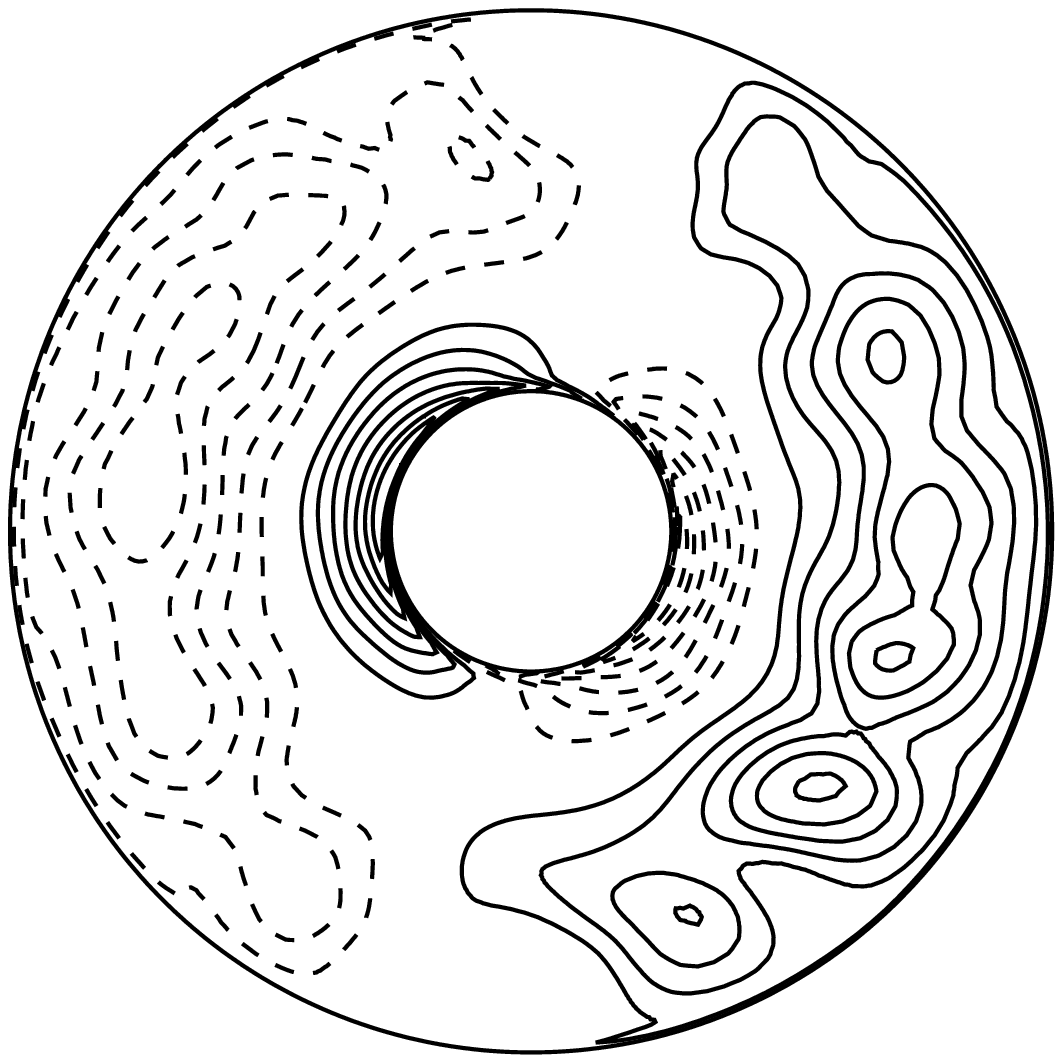}\\
            \vspace{-0.4cm}
            \textbf{(e2)}\\
            \includegraphics[width=\textwidth]{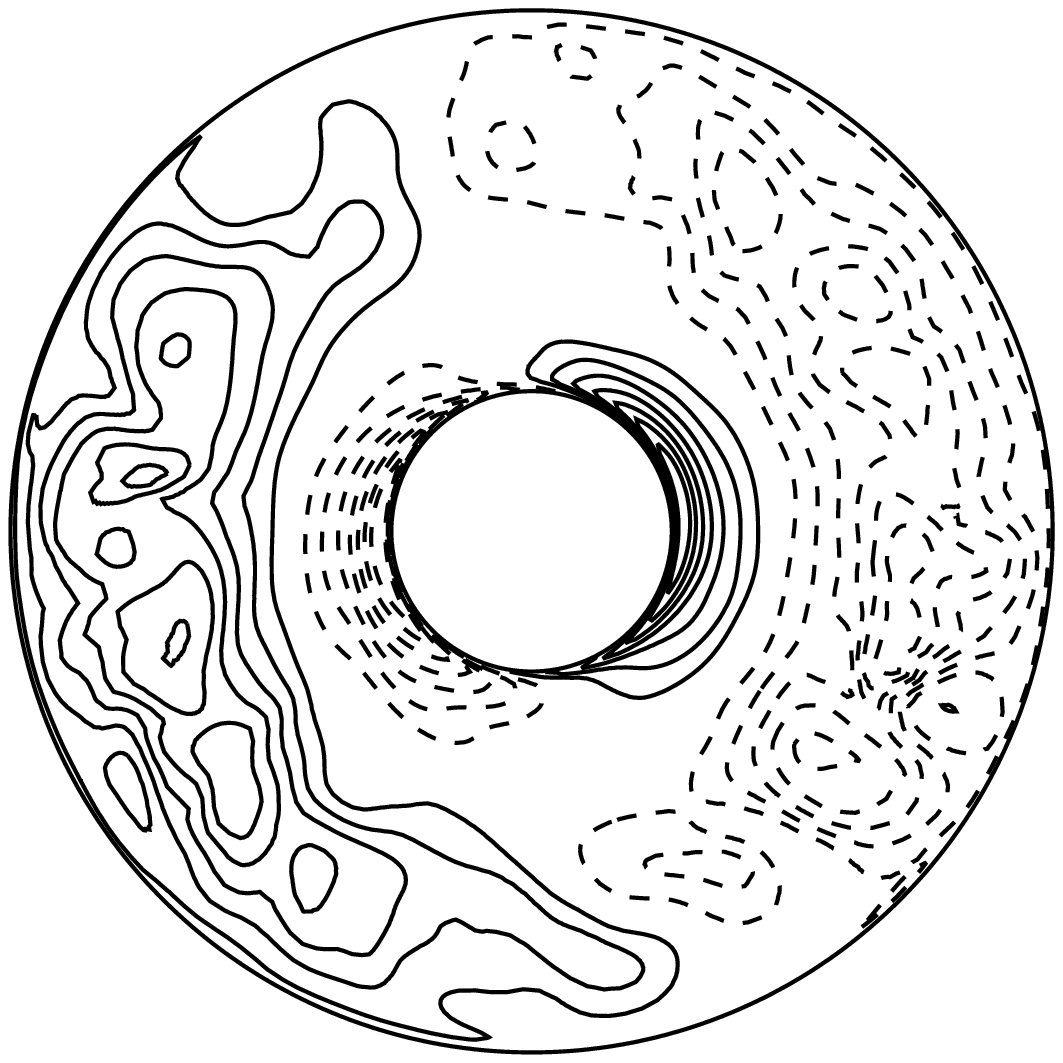}
        \end{minipage}
    \end{tabular}
    \caption{Contours of $u_z$ at $z=1/2$ plane for (a1,a2) $Po=0.01$; (b1,b2) $Po=0.02$; (c1,c2) $Po=0.0228$; (d1,d2) $Po=0.035$ and (e1,e2) $Po=0.05$ at two different instants: $t=0$ on the left and $t=300$ on the right. Solid lines are for positive $u_z$ and the dash lines for negative $u_z$. From top to bottom, the contour levels are different for different Poincar$\acute{\mathrm e}$ numbers.} \label{fig_uz}
\end{center}
\end{figure}

\begin{figure}
   \centerline{\includegraphics[width=\textwidth]{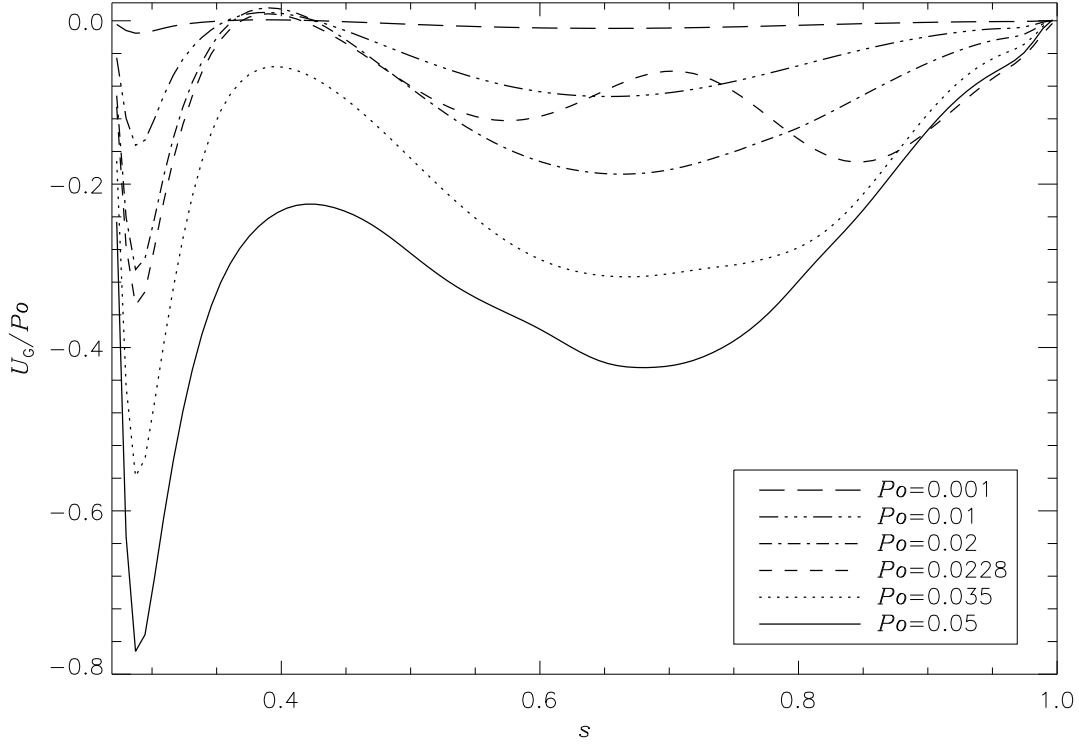}}
   \caption{Radial distribution of the scaled axisymmetric geostrophic flow $U_\mathrm{G}/Po$ for different $Po$ at $E=5\times10^{-5}$ in the precessing cylindrical annulus where $s$ is the radial coordinate. Note the velocity should vanish on the boundaries, but the discrete $u_\phi$ is not located right at $s=0.269$ and $s=1$ in the staggered-grid.}
   \label{figUG}
\end{figure}

To reveal the features of a nonlinear precessing flow, the total kinetic energy density $E_{\mathrm{kin}}(t)$ and the radial distribution of the geostrophic flow $U_\mathrm{G}(s)$ are introduced as
\begin{equation}
  E_{\mathrm{kin}}(t) = \frac{1}{2\pi(\Gamma^2-\Upsilon^2)}\int_0^1\int_{\Upsilon}^\Gamma\int_0^{2\pi}|\bm u(t)|^2s\mathrm{d}\phi\mathrm{d}s\mathrm{d}z,
\end{equation}
\begin{equation}
   U_\mathrm{G}(s)=\frac{1}{2\pi}\int_0^1 \int_0^{2\pi}\hat\phi\cdot\bm u(s,\phi,z)\mathrm{d}\phi\mathrm{d}z.
\end{equation}
 The results obtained from numerical simulations for different Poincar$\acute{\mathrm e}$ numbers are summarized with the evolution of $E_{\mathrm{kin}}$ in Fig.~\ref{figEkin}, the spatial structure of velocity component $u_z$ in Figure~\ref{fig_uz}, and the radial distribution of $U_\mathrm{G}$ in Figure~\ref{figUG}.

Due to viscous boundary effect, at small Ekman number $\sqrt{E}\ll 1$, the flow will take a long time $t\geq O(1/\sqrt{E})$ to reach nonlinear equilibrium.
The numerical simulation usually requests the computation run to the dimensionless time $t=1000\sim$3000, more than a hundred rotation periods. Fig.~\ref{figEkin} shows the evolution of kinetic energy density within the last period of 300 time units under nonlinear equilibrium. The start and end of this period are marked as $t=0$ and $t=300$. Note that the scalings of $E_{\mathrm{kin}}$ are different for each $Po$ in Fig.~\ref{figEkin}.

Several evident characteristics are revealed from the numerical simulations. First, as displayed in Fig.~\ref{figEkin}, the kinetic energy density is stable when the precession is weak ($Po\leq0.02$). In moderate precession, the kinetic energy increases and starts to change irregularly and nonperiodically. The magnitude of variation is also enhanced as $Po$ increases, but within a few percent.
Second, as illustrated in Fig. ~\ref{fig_uz}, the structure of the precessing flow, as expected, becomes disordered gradually when $Po$ increases. Comparing the left column at $t=0$ with the right column at $t=300$ of Figure~\ref{fig_uz}, the primary pattern of the spatial structure almost remains unchanged after hundreds of rotations for each $Po$ under nonlinear equilibrium. Third, the precessing flow consists of several waves (modes) that travels retrogradely (opposite to the direction of annulus rotation) and the predominant mode is $\bm u_{111}$ for $0.001\leq Po\leq 0.05$, in agreement with the result of experiments by Lin et al. (\cite{Lin2014}). The amplitude and frequency of the predominant mode is also shown in Table \ref{tab1}-\ref{tab2} which list the largest ten $|A_{mnk}|$ in the precessing flow for various $Po$.

In a \emph{cylinder}, when  the Ekman number $E$ is sufficiently small, the precessional force resonates directly with the cylindrical inertial mode whose eigen-frequency is close to the resonant value $1$ at any given aspect ratio $\Gamma$ (Poincar\'e mode), and the amplitude of the precessing flow $|\bm u|$ satisfies the asymptotic scaling law $|\bm u|=O(Po/\sqrt{E})$ at small $Po$, ~$|\bm u|=O(Po)$ at large $Po$ and the transition between them (Liao \& Zhang~\cite{Liao2012}).
In an \emph{annulus} of our simulation,  the nonlinear precessing flow, when $E$ is sufficiently small, is laminar with a constant amplitude at small $Po$ ($0<Po\leq0.02$) (Fig. \ref{figEkin} (a,b)), which obeys the scaling law $|\bm u|=O(Po/\sqrt{E})$. Thus we can define the weakly nonlinear regime as $0<Po/\sqrt{E}\leq C$ on the specific setup of our simulations, where $C=2.83$, a constant of $O(1)$.

The predominant mode is $\bm u_{111}$ in an \emph{annulus} of our setup ($\Gamma=1$), which is different from the case in a \emph{cylinder} where the mode $\bm u_{112}$ dominates when $\Gamma=1.045945$ (Liao \& Zhang~\cite{Liao2012}), even though the aspect ratios $\Gamma$ of both systems are almost the same. In a \emph{cylinder} with $\Gamma=1.045945$, the eigen-frequency of mode $\bm u_{112}$ is just 1, exactly the same as the precessional frequency ($\omega=1$), so the mode $\bm u_{112}$ would be resonated directly by Poincar$\acute{\mathrm e}$ forcing and dominates the precessing flow. However, in the \emph{annulus}, the largest three modes are $\bm u_{111}, \bm u_{113}$ and $\bm u_{112}$, which dominate the precessing flow. This is because their eigen-frequencies are 1.2748, 0.4968 and 0.7312, respectively (see Table \ref{tab1}-\ref{tab2}), in the vicinity of the frequency of Poincar\'e forcing ($\omega=1$). Other higher-order modes are strongly diminished by viscous dissipation, making small contributions to the precessing flow even though its frequency is close to the resonant value.

It is anticipated that the precessing flow has rich dynamics in annuli in the moderately precessing regime $C<Po/\sqrt{E}\leq 10$, since in \emph{cylinders}, the precessing flow is strongly turbulent where the inertial mode which is directly driven by Poincar$\acute{\mathrm e}$ force is still predominant, but many nonresonant inertial modes are spawned and their amplitudes have the same order as the primary mode. Figure \ref{figEkin} (d,e) and Figure \ref{fig_uz} (d,e) illustrate similar phenomenon for annulus as $Po$ increases, but the turbulence amplitude is not so high as in cylinders. The flow exhibits more stability in an annulus than in a cylinder because the newly excited inertial modes are reduced by the boundary effects with the inner sidewall involved.

When $Po/\sqrt{E}>10$,  on the contrary, the precessing flow has weak turbulence in \emph{cylinders}, the amplitude of kinetic energy variation is small and the spatial structure  is simpler than those of the turbulent flow when $1<Po/\sqrt{E}\leq10$ (Jiang et al.~\cite{Jiang2015}, Kong et al.~\cite{Kong2015}). For this reason, we will focused on the transition regime of $1<Po/\sqrt{E}\leq 10$ in our study for the geometry of \emph{annulus} where rich dynamics is expected.

From numerical simulations, it is found that the kinetic energy density of geostrophic flow is about 1\% of the total kinetic energy density $E_{\mathrm{kin}}$ when $Po=0.0228$, and increases to $10\%$ of $E_{\mathrm{kin}}$ when $Po=0.05$, which implies
 the geostrophic flow does not dominate the precessing flow when $Po\leq0.05$ or $Po/\sqrt{E}\leq7.07$. This is compatible with the numerical study in a \emph{cylinder} by Jiang et al. (\cite{Jiang2015}) where the geostrophic flow is also very weak when $1<Po/\sqrt{E}\leq 10$.

It has been proved that the nonlinear interaction of the inertial mode can not generate or maintain the geostrophic flow $U_\mathrm{G}(s)\hat\phi$ (Jiang et al.~\cite{Jiang2015}, Kong et al.~\cite{Kong2015}). In fact, the steady geostrophic flow is spawned by the nonlinear effect in the viscous boundary layers. The geostrophic flow is in the form of a differential rotation depending on $s$ and determined by the structure of the predominant inertial mode $\bm{u}_{111}$.  Figure~\ref{figUG} shows the geostrophic flow possesses a retrograde rotation with the lowest negative amplitude at $s=0.287$, and a prograde rotation  with a peak positive amplitude at $s=0.390$. When $Po>0.02375$, the prograde rotation will not survive and the geostrophic flow always rotates retrogradely. The numerical simulations also show that the geostrophic flow has an abnormal local amplitude peak at $s=0.7$ when $0.021875\leq Po\leq 0.03125$, which does not appear beyond this parametric range, i.e., $Po\leq 0.02$ or $Po\geq0.035$.  This feature of radial distribution has not been found in the precessing flow in \emph{cylinders} (Jiang et al.~\cite{Jiang2015}, Kong et al.~\cite{Kong2015}) and may result from the complicated nonlinear interaction as the inner boundary layer is involved. Thus the parameter regime $0.021875\leq Po\leq 0.03125$ could be regarded as the division between weak and strong nonlinearity in current study.

\subsection{Search for triadic resonance}
In order to check the existence of triadic resonance in the precessing flow in rotating annulus, the geometry of our numerical simulations is deliberately set to be the same as that of the laboratory experiment by Lin et al. (\cite{Lin2014}). The parameters adopted in their study are $Po$=(1.0E-2, ~1.0E-2, ~5.3E-3) in correspondence with $E$=(1.0E-5, ~1.0E-5, ~5.0E-6) that lead the velocity scaling as $Po/\sqrt{E}$=(3.16, ~4.43, ~2.37), ranging in the transition parameter regime $1<Po/\sqrt{E}\leq 10$.
In our simulations, several Poincar$\acute{\mathrm e}$ numbers are set from $Po=0.001$ to 0.05, leading $Po/\sqrt{E}$ in the range (0.141, ~7.07) which is compatible with the values of the laboratory experiment.

Suppose triadic resonance is the mechanism by which the laminar flow breaks and transits to the disordered flow in a precessing annulus. Then there must exist three  inertial modes with large amplitude that dominate the flow and obey the triadic resonant conditions (\ref{triadic}).
To check the existence of triadic resonance, an effective way is to use the general representation (\ref{eqn_decomp}) to calculate the coefficients $A_{mnk}$ for each mode of the precessing flow and see if the wave number $m, n$ and the corresponding frequency $\omega_{mnk}$ of the large modes satisfy the conditions.  For the modes with $m\geq 1, n\geq 1$ and $k\geq 1$, $A_{mnk}$ can be derived from an integration of (\ref{eqn_decomp})  with the numerical solution $\bm u$ as
\begin{equation}\label{eqn_amnk}
    A_{mnk}(t)=2\int_0^1\int_{\Upsilon}^{\Gamma}\int_0^{2\pi}\left(\bm u_{mnk}^*\cdot\bm u\right)s\mathrm{d}\phi \mathrm{d}s\mathrm{d}z,
\end{equation}
where $\bm u_{mnk}^*$ refers to the complex conjugate of mode $\bm u_{mnk}$ (eigenmode, see Greenspan~\cite{Greenspan1968}).
For other modes with $m=0$ or $n=0$, ~$A_{mnk}$ can be calculated in the same way. It should be noted that $\bm u_{mnk}^*$ and $\bm u_{mnk}$ are orthogonal and  normalized (Zhang \& Liao~\cite{ZL2017}).

The values of $|A_{mnk}|$ calculated from our numerical results are listed in Table~\ref{tab1}-\ref{tab2} for different Poincar$\acute{\mathrm e}$ numbers. Through extensive search in the transitional regime from laminar flow to disordered flow when $0.0228\leq Po\leq 0.05$ (no triadic resonance is expected taking place outside of transitional regime), there is no indication that the triadic resonant exists with three inertial modes dominating the precessing flow and satisfying the conditions (\ref{triadic}).
  \begin{table}
    \bc
    \begin{minipage}[]{120mm}
    \caption[]{Top ten coefficients $|A_{mnk}|$ for $Po=0.01, 0.02$ and 0.0228. \label{tab1}}\end{minipage}
    \setlength{\tabcolsep}{2.5pt}
    \small
 \begin{tabular*}{12cm}{@{\extracolsep{\fill}}cccc|cc}
  \hline
$(m,n,k)$ & $\omega_{mnk}$ & $|A_{mnk}|_{Po=0.01}$ & $|A_{mnk}|_{Po=0.02}$ & $|A_{mnk}|_{Po=0.0228}$ & $(m,n,k)$\\
  \hline
  ~(1,1,1)\phantom{$^{-}$} &  \phantom{-}1.2748  & 0.7373E-02 & 0.1468E-01 & 0.1668E-01 & ~(1,1,1)\phantom{$^{-}$}\\
  ~(1,1,3)\phantom{$^{-}$} &  \phantom{-}0.4968  & 0.2992E-02 & 0.5643E-02 & 0.6806E-02 & ~(1,1,3)\phantom{$^{-}$}\\
  ~(1,1,2)\phantom{$^{-}$} &  \phantom{-}0.7312  & 0.2814E-02 & 0.5558E-02 & 0.6798E-02 & ~(1,1,2)\phantom{$^{-}$}\\
  ~(1,1,5)\phantom{$^{-}$} &  \phantom{-}0.2986  & 0.8035E-03 & 0.1524E-02 & 0.1535E-02 & ~(1,1,5)\phantom{$^{-}$}\\
  ~(1,1,1)$^{-}$           &   -1.0008           & 0.5179E-03 & 0.9934E-03 & 0.1205E-02 & ~(1,1,1)$^{-}$          \\
  ~(1,3,3)\phantom{$^{-}$} &  \phantom{-}1.1938  & 0.4888E-03 & 0.7894E-03 & 0.1103E-02 & ~(1,3,3)\phantom{$^{-}$}\\
  ~(1,3,5)\phantom{$^{-}$} &  \phantom{-}0.8118  & 0.3819E-03 & 0.7258E-03 & 0.9145E-03 & ~(1,3,4)\phantom{$^{-}$}\\
  ~(1,3,4)\phantom{$^{-}$} &  \phantom{-}0.9722  & 0.3796E-03 & 0.7256E-03 & 0.7418E-03 & ~(1,1,3)$^{-}$\\
  ~(1,1,3)$^{-}$           &   -0.4439           & 0.3159E-03 & 0.6754E-03 & 0.7358E-03 & ~(1,3,5)\phantom{$^{-}$} \\
  ~(1,1,9)\phantom{$^{-}$} &  \phantom{-}0.1650  & 0.3004E-03 & 0.6032E-03 & 0.6817E-03 & ~(1,1,9)\phantom{$^{-}$}\\
\hline
\end{tabular*}
\ec
\tablecomments{12cm}{The superscript minus sign of  $(m,n,k)^{-}$ refers to the corresponding mode $\bm u_{mnk}$ with a negative angular frequency $\omega<0$. The order of modes according to $|A_{mnk}|$ for $Po=0.0228$ is slightly different from that for $Po=0.01$, 0.02, and the frequency can be found in the second column correspondingly.}
\end{table}
\begin{table}
    \bc
    \begin{minipage}[]{120mm}
    \caption[]{Top ten coefficients $|A_{mnk}|$ for $Po=0.035$ and $Po=0.05$. \label{tab2}}\end{minipage}
    \setlength{\tabcolsep}{2.5pt}
    \small
 \begin{tabular*}{12cm}{@{\extracolsep{\fill}}ccc|ccc}
  \hline
$(m,n,k)$ & $\omega_{mnk}$ & $|A_{mnk}|_{Po=0.035}$ & $|A_{mnk}|_{Po=0.05}$  & $\omega_{mnk}$  & $(m,n,k)$\\
  \hline
~(1,1,1)\phantom{$^{-}$} & \phantom{-}1.2748  & 0.2558E-01 & 0.3536E-01   &  \phantom{-}1.2748 & ~(1,1,1)\phantom{$^{-}$}\\
~(1,1,2)\phantom{$^{-}$} & \phantom{-}0.7312  & 0.1077E-01 & 0.1799E-01   &  \phantom{-}0.7312 & ~(1,1,2)\phantom{$^{-}$}\\
~(1,1,3)\phantom{$^{-}$} & \phantom{-}0.4968  & 0.8420E-02 & 0.1104E-01   &  \phantom{-}0.4968 & ~(1,1,3)\phantom{$^{-}$}\\
~(1,1,5)\phantom{$^{-}$} & \phantom{-}0.2986  & 0.2295E-02 & 0.4926E-02   &  \phantom{-}0.9736 & ~(3,1,1)\phantom{$^{-}$}\\
~(9,1,2)$^{-}$           & -0.3520            & 0.1464E-02 & 0.2807E-02   &  -1.0008           & ~(1,1,1)$^{-}$ \\
~(1,3,3)\phantom{$^{-}$} & \phantom{-}1.1938  & 0.1372E-02 & 0.2541E-02   &  \phantom{-}0.2986 & ~(1,1,5)\phantom{$^{-}$}\\
~(1,1,1)$^{-}$           & -1.0008            & 0.1300E-02 & 0.2345E-02   &  \phantom{-}1.1938 & ~(1,3,3)\phantom{$^{-}$}\\
~(1,1,3)$^{-}$           & -0.4436            & 0.1256E-02 & 0.2144E-02   &  \phantom{-}1.6384 & ~(2,2,1)\phantom{$^{-}$}\\
~(1,3,4)\phantom{$^{-}$} & \phantom{-}0.9722  & 0.1237E-02 & 0.2114E-02   &  \phantom{-}0.6838 & ~(0,2,4)\phantom{$^{-}$}\\
~(1,1,9)\phantom{$^{-}$} & \phantom{-}0.1650  & 0.1095E-02 & 0.2032E-02   &  \phantom{-}0.8254 & ~(4,1,1)\phantom{$^{-}$}\\
\hline
\end{tabular*}
\ec
\end{table}

Table~\ref{tab1} gives the largest ten coefficients $|A_{mnk}|$ for $Po=0.01, ~0.02$ and 0.0228 with $Po/\sqrt{E}=1.41, ~2.83$ and 3.23, respectively. The top three modes are $\bm u_{111}$, $\bm u_{113}$ and $\bm u_{112}$ for $Po=0.0228$ which shows no triadic resonance exists.
When $Po$ increases further to 0.035 and 0.05 with $Po/\sqrt{E}$ ranging from 4.95 to 7.07, the precessing flow undergoes stronger nonlinear interactions and the top ten coefficients $|A_{mnk}|$ are presented in Table~\ref{tab2}. The dominant modes are $\bm u_{111}$, $\bm u_{112}$ and $\bm u_{113}$, almost the same as the case for smaller $Po$, except that $|A_{112}|$ surpasses $|A_{113}|$. The lack of three modes that satisfy the conditions (\ref{triadic}) also does not indicate the existence of triadic resonance.

\section{Summary and discussion}
\label{sect:discuss}
We have studied the precession-driven flow, which is bounded in a cylindrical annulus by numerical simulations with a 3-D finite difference method.
A parallel computation code is developed to solve the nonlinear fluid motion and the flow features are elucidated with the evolution of kinetic energy density, spatial structure of $u_z$ and the radial distribution of geostrophic flow.
In current study, we intentionally set $\Gamma=1$ and $\Upsilon=0.269$ in the annular geometry in order to compare with the results of laboratory experiments.
By decomposing the nonlinear solution into a complete inertial-mode set, we do not find the well-known triadic resonance existing in the precessing flow, which is inconsistent with the findings in laboratory experiments by Lin et al.~(\cite{Lin2014}), but in agreement with the results by Kong et al.~(\cite{Kong2015}) and Jiang et al.~(\cite{Jiang2015}).

To offer deep insight into understanding of the precessional flow in an annulus,
we suggest an extensive experimental study on that the Poincar\'e force resonates directly
with the basic inertial modes by setting the inner radius $r_\mathrm{i}$ and
outer radius $r_\mathrm{o}$ with fixed $h$ to satisfy the resonance condition:
$$
\sigma_{1nk}^2=\frac{(n\pi)^2}{(n\pi)^2+\xi_{1nk}^2}=\left(\frac{1}{2}\right)^2,
~~(n=1,3,5,\cdots,~k=0,1,2,3,\cdots)
$$
where $\sigma$ is the half-frequency of inertial mode $u_{1nk}$ and the radial wave number $\xi$ is determined by the following equation:
\begin{eqnarray*}
&&\left[\xi r_\mathrm{i} \mathrm{J}_0(\xi r_\mathrm{i} )+
(1/\sigma-1)\mathrm{J}_1(\xi r_\mathrm{i})
\right]
\left[\xi r_\mathrm{o} \mathrm{Y}_0(\xi r_\mathrm{o} )+
(1/\sigma-1)\mathrm{Y}_1(\xi r_\mathrm{o})
\right]\\
&=&
\left[\xi r_\mathrm{i} \mathrm{Y}_0(\xi r_\mathrm{i} )+
(1/\sigma-1)\mathrm{Y}_1(\xi r_\mathrm{i})
\right]
\left[\xi r_\mathrm{o} \mathrm{J}_0(\xi r_\mathrm{o} )+
(1/\sigma-1)\mathrm{J}_1(\xi r_\mathrm{o})
\right],
\end{eqnarray*}
in which $\mathrm{J}_m$ and $\mathrm{Y}_m$ are standard Bessel functions of the first and second kind, respectively; the subscripts of $\xi$ and $\sigma$ are omitted.

 Our simulations demonstrate that the inertial mode $\bm u_{111}$ always dominates the precessing flow when $0.001\leq Po\leq 0.05$.  The flow is stable with a constant amplitude of kinetic energy when $Po\leq0.02$. As the precession rate increases to the range $0.02<Po\leq 0.05$, the laminar flow breaks and shifts into the transition regime to disordered flow where the nonlinear viscous boundary effects play an important role in the dynamics of precessing flow instead of the mechanism of triadic resonance. In an annulus, the flow is more stable than in a cylinder as an additional inner boundary restricts the growth of inertial modes with viscous damping. The geostrophic flow is weak when $Po<0.05$ and its radial distribution shows that there is a boundary parameter region $0.02185\leq Po\leq 0.03125$ between the weak and strong nonlinearity of the precessional flow.

Since the size of the Ekman number in the Earth's outer core is extremely small and the precessing flow in the outer core is likely to be highly turbulent, it is desirable to expand our knowledge on this problem to smaller Ekman numbers and larger Poincar\'e numbers, which makes the parallel computation very expensive, as a smaller time step and a higher spatial resolution are required. A new code of parallel computation to fulfill the task based on many-core architecture (such as Intel MIC) is under construction instead of the present one built on multi-core CPUs. Moreover, a precession-driven planetary dynamo model can also be carried out based on this numerical framework.

\begin{acknowledgements}
This work was supported by NSFC under No. 11673052, 41661164034 and CAS Pilot Project under No. XDB18010203.
\end{acknowledgements}

\label{lastpage}
\end{document}